\documentclass[12pt]{article}
\usepackage{pdproc,epsfig}

\begin{document}

\renewcommand{\thefootnote}{\alph{footnote}}

\begin{flushright}
RAL-TR-2004-004 \\
hep-ph/0402006 \\
31 Jan 2004
\end{flushright}

\title{
STATUS OF TRI/BI-MAXIMAL NEUTRINO MIXING~\protect\footnote{
Written version of a talk presented by W. G. Scott
at the ``Second Workshop on Neutrino Oscillations in Venice'' 
(NOVE-II), Venice, Italy. 3-5th Dec.\ 2003.}} \vspace{-2mm}

\author{P. F. HARRISON}
\address{
Physics Department, University of Warwick, UK. \\
 {\rm E-mail: p.f.harrison@warwick.ac.uk}}

  \centerline{\footnotesize and}

\author{W. G. SCOTT}

\address{Centre for Fundamental Physics (CFFP), 
and Particle Physics Department (PPD), \\
CCLRC Rutherford Appleton Laboratory, 
Chilton, Didcot, OX11-0QX, UK\\
 {\rm E-mail: w.g.scott@rl.ac.uk}} \vspace{-7mm}

\abstract{
Tri/bi-maximal mixing (TBM) is a specific lepton mixing ansatz,
which describes the trend of the current neutrino oscillation data, 
in particular the recent SNO and KAMLAND results.
The significant feature of TBM in this respect is
$|U_{e2}|^2=|U_{\mu 2}|^2=|U_{\tau 2}|^2=1/3$,
and we say that the $\nu_2$ is {\em tri}-maximally mixed.
We have generalised the TBM ansatz 
to a generic mixing matrix with the $\nu_2$ trimaximally mixed,
whereby the neutrino mass matrix in the lepton flavour basis
takes the form of a general S3 group matrix 
($3 \times 3$ `magic-square').
In exact TBM the charged-lepton mass matrix
in the neutrino mass basis 
(where the neutrino mass matrix is diagonal) 
takes the form of a general S3 class operator.
The neutrino mass matrix in the flavour basis
is a particular S3 group matrix which is also 
an $S1 \subset S2 \subset S3$ group-chain class operator,
whereby the neutrino mass eigenstates are distinguished 
by their `mutativity' ($M_i=\pm 1$) and `democracy' ($D_i=0,3$)
which are both good quantum numbers in exact TBM.} \vspace{-2mm}
   
\normalsize\baselineskip=15pt

\section{Tri/Bi-Maximal Mixing} \vspace{-1mm}
  Tri/bi-maximal mixing (TBM) is a specific lepton mixing ansatz,
which is now already five years old in the literature \cite{tbm}: \vspace{-2mm}
\begin{eqnarray}
     \matrix{  \hspace{0.4cm} \nu_1 \hspace{0.3cm}
               & \hspace{0.2cm} \nu_2 \hspace{0.3cm}
               & \hspace{0.2cm} \nu_3  \hspace{0.4cm} }
                                      \hspace{0.2cm} \nonumber \\
\hspace{1.0cm}U \hspace{0.3cm} = \hspace{0.3cm}
\matrix{ e \hspace{0.2cm} \cr
         \mu \hspace{0.2cm} \cr
         \tau \hspace{0.2cm} }
\left( \matrix{ \sqrt{\frac{2}{3}}  &
                     \frac{1}{\sqrt{3}} &
                                   0  \cr
 -\frac{1}{\sqrt{6}}  &
          \frac{1}{\sqrt{3}} &
 -\frac{1}{\sqrt{2}}   \cr
      \hspace{2mm}
 -\frac{1}{\sqrt{6}} \hspace{2mm} &
         \hspace{2mm}
            \frac{1}{\sqrt{3}} \hspace{2mm} &
                                 \frac{1}{\sqrt{2}}
                                       \hspace{2mm} \cr } \right)
\label{tb0}
\end{eqnarray}
In TBM (Eq.~\ref{tb0})
the intermediate mass neutrino $\nu_2$ is {\em tri}-maximally mixed
between all three lepton flavours,
as in the original trimaximal scheme \cite{tri},
while the heavy neutrino $\nu_3$ is {\em bi}-maximally mixed
between mu and tau flavours,
as in the original bimaximal scheme \cite{bim}
(where we take a conventional mass-hierarchy for simplicity). 
TBM may also be viewed as a special case
of the Altarelli-Feruglio mixing scheme \cite{alf}.

Of course we recognise that TBM (Eq.~\ref{tb0})
is unlikely to be realised {\em exactly} in nature~\cite{syms}.
In particular, the absence of $CP$-violation 
predicted in exact TBM would be surprising,
given the success of the Standard Model explanation 
of $CP$-violation in the case of the quarks.
We see Eq.~\ref{tb0} as a useful summary 
of the trends in the current experimental data, 
suggestive of possible underlying symmetries,
but really only a step towards a deeper understanding
of masses and mixings in general.
It should not be forgotten, however, that
with $|U_{e2}|^2=|U_{\mu 2}|^2=|U_{\tau 2}|^2=1/3$ in TBM, 
just as in the original trimaximal scheme \cite{tri},
there is a legitimate claim to a degree of predictivity (see Section~2),
supporting our basic approach
(which was in fact initially focussed on the {\em quarks} \cite{hs1},
with no regard to any neutrino oscillation data).

\section{The Experimental Evidence}
\noindent The evidence for $\nu_3$ 
being bimaximally mixed (at least approximately)  
comes from the atmospheric neutrino data~\cite{atm}
supported by the K2K data~\cite{k2k}, and from the reactor data 
(especially CHOOZ~\cite{choz}  and PALO VERDE~\cite{palo}).
The key experimental numbers are: 
$|U_{\mu 3}|^2 \simeq 0.50 \pm 0.11$
and $|U_{e3}|^2 < 0.04$ (90\% CL)\cite{fogl},
clearly consistent with TBM: 
\begin{eqnarray}
     \matrix{  \hspace{0.1cm} \nu_1 \hspace{0.2cm}
               & \hspace{0.4cm} \nu_2 \hspace{0.2cm}
               & \hspace{0.4cm} \nu_3  \hspace{0.2cm} }
                                      \hspace{0.4cm} \nonumber \\
(|U_{l \nu}|^2) \hspace{2mm} = \hspace{2mm}
\matrix{ e \hspace{0.2cm} \cr
         \mu \hspace{0.2cm} \cr
         \tau \hspace{0.2cm} }
\left( \matrix{ 2/3  &
                      1/3 &
                              0 \cr
                1/6 &
                    1/3 &
                             1/2  \cr
      \hspace{2mm} 1/6 \hspace{2mm} &
         \hspace{2mm}  1/3 \hspace{2mm} &
           \hspace{2mm} 1/2  \hspace{2mm} \cr } \right)
\label{obx}
\end{eqnarray}
where we represent the mixing 
in terms of the moduli-squared 
of the mixing elements.

We focus here on the evidence for the $\nu_2$ 
being trimaximally mixed.
Table~1 gives \vspace{-3mm}
\begin{table}[h] \vspace{-2mm}
\caption{The SNO results for 2002 (pure $D_2$O) 
and 2003 ($D_2$O + Salt), 
assuming an undisorted $^8$B spectrum.
The naive average given is the average $CC/NC$ ratio
assuming uncorrelated systematics.}
\label{tab:exp} \vspace{2mm}
  \small
  \begin{tabular}{||c|c|c|c||}\hline\hline
  {} &{} &{} &{}\\
 SNO results  &  CC / $10^{ 6}$ cm$^{-2}$ s$^{-1}$ 
   &  NC / $10^{ 6}$ cm$^{-2}$ s$^{-1}$ 
   & CC/NC Ratio\\
  {} &{} &{} &{}\\
  \hline
  {} &{} &{} &{}\\
  Pure D$_2$O & $1.76 \pm
   \raisebox{0.95ex}{0.06} \raisebox{-0.95ex}{\hspace{-6.7mm}0.05}
\pm 0.09$ 
              & $5.09 \pm 
  \raisebox{0.95ex}{0.44} \raisebox{-0.95ex}{\hspace{-6.7mm}0.43}  
\pm \raisebox{0.95ex}{0.46} \raisebox{-0.95ex}{\hspace{-6.7mm}0.43}
$ & \hspace{2mm}
  $0.346 \pm 
\raisebox{0.95ex}{0.032} \raisebox{-0.95ex}{\hspace{-8.5mm}0.031} 
\pm \raisebox{0.95ex}{0.036} \raisebox{-0.95ex}{\hspace{-8.5mm}0.034}
$ \hspace{2mm} \\
  {} &{} &{} &{}\\
  \hline
  {} &{} &{} &{}\\
  D$_2$O + Salt &  $1.70 \pm 0.07 \pm 
\raisebox{0.95ex}{0.09} \raisebox{-0.95ex}{\hspace{-6.7mm}0.10}
$  & $4.90 \pm 0.24 \pm
\raisebox{0.95ex}{0.29} \raisebox{-0.95ex}{\hspace{-6.7mm}0.27}
$ & \hspace{2mm}
  $0.347 \pm 0.022 \pm 0.028$ \hspace{2mm} \\ 
  {} &{} &{} &{}\\
\hline
  {} &{} &{} &{}\\
  Naive Avg. & $-$ & $-$ &
  $0.35 \pm 0.03$ \\
  {} &{} &{} &{}\\
  \hline\hline
\end{tabular} 
\end{table}

\noindent the measured charged-current ($CC$) 
and neutral-current ($NC$) fluxes in SNO,
for both the 2002 \cite{sn02} and 2003 \cite{sn03} analyses, 
corresponding to pure D$_2$O and
D$_2$O + Salt running respectively.
In the LMA solution to the solar neutrino problem 
(Appendix), spectral distortion
over the SNO energy range is expected to be small,
and we quote the results for an undistorted $^8$B spectrum,
which should be statistically the most precise.
The naive average of the $CC/NC$ ratios in Table~1  
is the average  of the pure D$_2$O and the D$_2$O + Salt ratios
assuming (optimistically) {\em un}correlated systematics.

The CC/NC ratio in SNO
measures directly the electron survival probability 
in the base of the LMA  `bathtub',
which gives $|U_{e2}|^2 \simeq 0.35 \pm 0.03$ directly, 
see Figure~1. 

\begin{figure}[h]
\begin{center}
\epsfig{figure=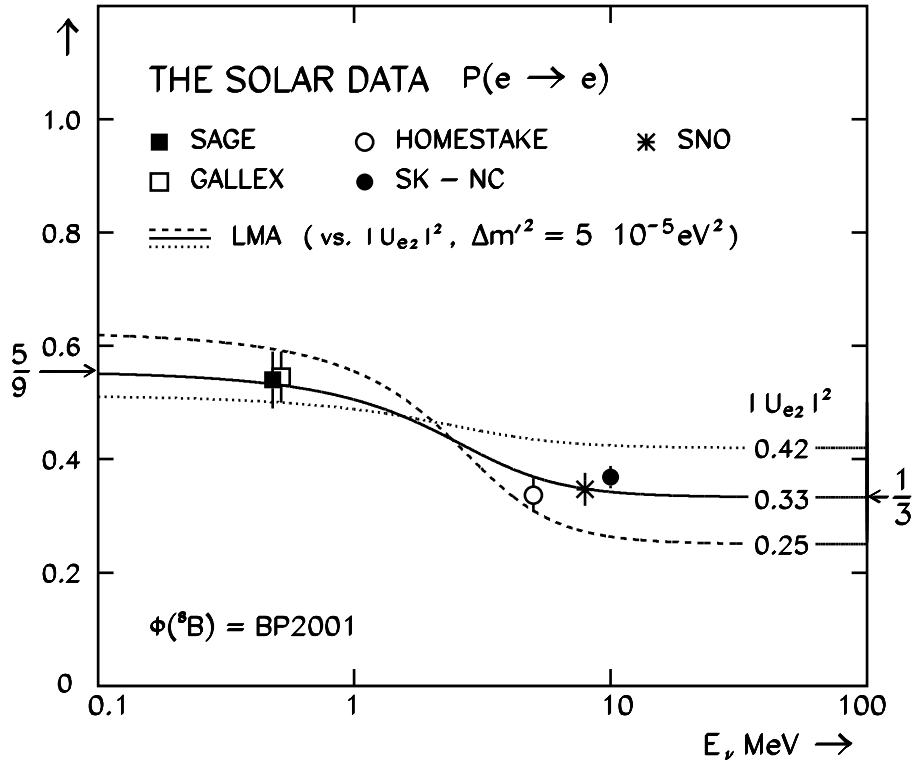,width=12.0cm}
\end{center}
\caption{The solar neutrino data plotted as a function 
of neutrino energy. 
The SNO point is the average $CC/NC$ ratio
assuming no $^8$B spectral distortion,
while the other experiments are plotted
assuming BP2001 fluxes.
The solid curve is the `$5/9-1/3-5/9$' bathtub prediction
(first plotted in our 1996 paper~\protect\cite{hps2} 
in the context of trimaximal mixing) 
showing that the $\nu_2$ is trimaximally mixed.}
\label{fig:sol}
\end{figure}
\noindent This is certainly consistent with TBM, for which $|U_{e2}|^2=1/3$.
The other experiments in Figure~1 
are dependent on the BP2001 fluxes \cite{bp01}. 
At low energies, vacuum mixing predominates
and the gallium experiments~\cite{gall} are consistent with
the corresponding vacuum prediction 
$<P(e \rightarrow e)>$ $\rightarrow$ $|U_{e1}|^2+|U_{e2}|^"+|U_{e3}|^2=5/9$.
Note that the far high energy end of the `bathtub',
corresponding to the breakdown of the adiabatic approximation,
plays no role in the LMA solution.

Before leaving the solar data, 
we pause breifly to make a slightly immodest, 
but we believe justifiable claim,
to a degree of predictivty in our overall approach.
While the original trimaximal ansatz was wrong,
in that it failed to predict small $|U_{e3}|$,
it {\em did} succed in predicting 
$|U_{e1}|^2=|U_{\mu 2}|^2=|U_{\tau 2}|^2=1/3$,
ie.\ {\em two} independent (Extended) Standard Model parameters,
at a time when many theorists were backing 
small angles at the solar scale (the SMA solution).
In particular, the `$5/9-1/3-5/9$' bathtub,
characteristic of a trimaximally mixed solar neutrino
was plotted out in the context of trimaximal mixing,
in our 1996 paper~\cite{hps2}, 
more than half a decade before the SNO data,
and we have taken the liberty of reproducing 
that Figure in an Appendix.
Despite the updated context, 
that curve (Figure~3) 
clearly fits the present-day data rather well (Figure~1),
strongly suggesting 
that we may have been on the right track.

\begin{figure}[h]
\begin{center}
\epsfig{figure=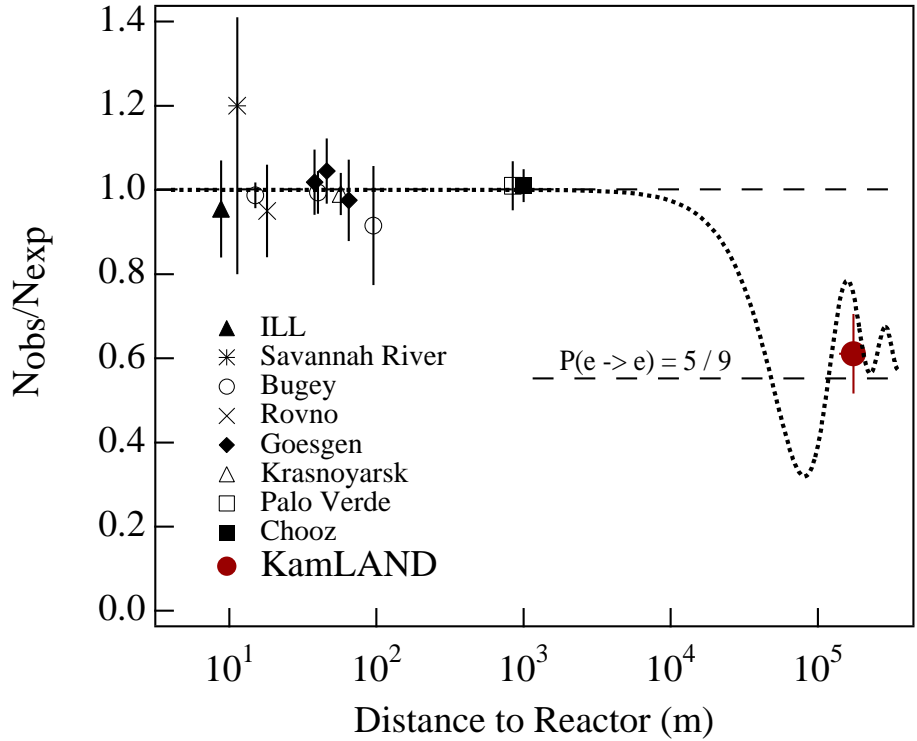,width=13.0cm}
\end{center}
\caption{The reactor data 
and in particular the KAMLAND result~\protect\cite{kaml}.
The data are consistent with the tri/bi-maximal vacuum prediction
$<P(e \rightarrow e)>$ $\rightarrow$ 
$|U_{e1}|^2+|U_{e2}|^2+|U_{e3}|^2=5/9$.~\protect\cite{tbm}}
\label{fig:kam}
\end{figure}

Figure~2 shows the reactor data 
and in particular the KAMLAND result,
which spectacularly confirms the LMA solution
to the solar neutrino problem. 
The measured average supression in KAMLAND
$<P(e \rightarrow e)>$ $\simeq$ $0.61 \pm 0.09$
is clearly consistent with the tri/bi-maximal prediction
$<P(e \rightarrow e)>$ $\rightarrow$ $|U_{e1}|^2+|U_{e2}|^2+|U_{e3}|^2=5/9$~\cite{tbm}. 
  
\section{The Generalisation to `S3 Group Mixing'}
     As concerns the lepton mass matrices,
it should be made clear that, for general complex mixing, 
the so-called flavour basis 
(defined by the charged-lepton mass matrix being diagonal) 
is not unique.
There remains the freedom to redefine the phases 
of the charged-lepton mass eigenstates,
which in turn modifies the phases of the off-diagonal elements
of the neutrino mass matrix.
This freedom 
may be removed by requiring 
that the imaginary part of the neutrino mass matrix
is proportional to~\cite{muta} 
$\epsilon_{\alpha \beta} := \epsilon_{\alpha \beta \gamma}$
($\gamma \neq \alpha , \beta$) $\alpha , \beta , \gamma = 1,2,3$.
This particlar flavour basis will be referred to as the `epsilon basis'.
Note that the epsilon matrix ($\epsilon_{\alpha \beta}$)
is circulant \cite{hs1}. 

One way to define 
the original trimaximal ansatz is to require
that the neutrino mass matrix (hermitian-square) 
in the lepton flavour basis be circulant: 
\begin{eqnarray} 
     \matrix{  \hspace{0.1cm} e \hspace{0.2cm}
               & \hspace{0.2cm} \mu \hspace{0.2cm}
               & \hspace{0.2cm} \tau  \hspace{0.2cm} }
                                      \hspace{0.2cm} \nonumber \\
M_{\nu}^2 \hspace{2mm} = \hspace{2mm}
\matrix{ e \hspace{0.2cm} \cr
         \mu \hspace{0.2cm} \cr
         \tau \hspace{0.2cm} }
\left( \matrix{ a  &
                      b &
                              \bar{b} \cr
                \bar{b} &
                    a &
                             b  \cr
      \hspace{2mm} b \hspace{2mm} &
         \hspace{2mm}  \bar{b} \hspace{2mm} &
           \hspace{2mm} a  \hspace{2mm} \cr } \right)
\label{mc3}
\end{eqnarray}
where $\bar{b}$ denotes the complex conjugate of $b$.
By definition a $3 \times 3$ circulant matrix 
is a general C3 group matrix 
(ie.\ a linear combination of C3 group elements ) 
in the natural representation of C3, and  
we are automatically in the epsilon basis.
    
We have generalised this construction~\cite{char}
to S3 group matrices, yielding the form:
\begin{eqnarray}
\matrix{  \hspace{0.1cm} e \hspace{0.2cm}
               & \hspace{0.2cm} \mu \hspace{0.2cm}
               & \hspace{0.2cm} \tau  \hspace{0.2cm} }
                                      \hspace{2.2cm}
     \matrix{  \hspace{0.1cm} e \hspace{0.2cm}
               & \hspace{0.2cm} \mu \hspace{0.2cm}
               & \hspace{0.2cm} \tau  \hspace{0.2cm} }
                                      \hspace{0.2cm} \nonumber \\
M_{\nu}^2 \hspace{2mm} = \hspace{2mm}
\matrix{ e \hspace{0.2cm} \cr
         \mu \hspace{0.2cm} \cr
         \tau \hspace{0.2cm} }
\left( \matrix{ a  &
                      b &
                              \bar{b} \cr
                \bar{b} &
                    a &
                             b  \cr
      \hspace{2mm} b \hspace{2mm} &
         \hspace{2mm}  \bar{b} \hspace{2mm} &
           \hspace{2mm} a  \hspace{2mm} \cr } \right)
+
 \hspace{2mm}
\matrix{ e \hspace{0.2cm} \cr
         \mu \hspace{0.2cm} \cr
         \tau \hspace{0.2cm} }
\left( \matrix{ x  &
                      z &
                              y \cr
                z &
                    y &
                             x  \cr
      \hspace{2mm} y \hspace{2mm} &
         \hspace{2mm}  x \hspace{2mm} &
           \hspace{2mm} z  \hspace{2mm} \cr } \right)
\label{ms3}
\end{eqnarray}
which is a circulant plus a retrocirculant 
or a general (hermitian) $3 \times 3$ `magic square'.
Again we are automatically in the epsilon basis.
It is of course the magic square property
(all row/column-sums equal) which guarantees
one trimaximal eigenvector:
\begin{eqnarray}
     \matrix{  \hspace{0.1cm} \nu_1 \hspace{0.6cm}
               & \hspace{0.6cm} \nu_2 \hspace{0.2cm}
               & \hspace{0.4cm} \nu_3  \hspace{0.3cm} }
                                      \hspace{0.4cm} \nonumber \\
(|U_{l \nu}|^2) \hspace{2mm} = \hspace{2mm}
\matrix{ e \hspace{0.2cm} \cr
         \mu \hspace{0.2cm} \cr
         \tau \hspace{0.2cm} }
\left( \matrix{ 2/3-|X|^2  &
                      1/3 &
                              |X|^2 \cr
                2/3-|Y|^2&
                    1/3 &
                              |Y|^2  \cr
      \hspace{2mm} 2/3-|Z|^2 \hspace{2mm} &
         \hspace{2mm}  1/3 \hspace{2mm} &
           \hspace{2mm} |Z|^2  \hspace{2mm} \cr } \right)
\label{s3g}
\end{eqnarray}
where we identify neutrino masses $m_1 < m_2 < m_3$ 
and mixing angles $\chi$,$\phi$ as follows: \vspace{1mm}
\begin{eqnarray}
a=\frac{2}{3} m_1^2+\frac{m_2^2}{3}; 
  \hspace{25mm} b=\frac{m_1^2}{6}+\frac{m_2^2}{3}-\frac{m_3^2}{2}
                    +i\frac{m_3^2-m_1^2}{2\sqrt{3}}\sin 2\chi; \hspace{1.4cm} \\
x=(m_3^2-m_1^2) |X|^2; \hspace{17mm}
      |X|^2=\frac{1}{3}-\frac{1}{3}\cos 2 \chi \cos 2 \phi; \hspace{3.7cm}\\
y=(m_3^2-m_1^2) |Y|^2; \hspace{17mm}
      |Y|^2=\frac{1}{3}+\frac{\cos 2 \chi \cos 2 \phi}{6} 
                 -\frac{1}{2 \sqrt{3}}\cos 2 \chi \sin 2 \phi; \hspace{0.4cm} \\
z=(m_3^2-m_1^2) |Z|^2; \hspace{17mm}
      |Z|^2=\frac{1}{3}+\frac{\cos 2 \chi \cos 2 \phi}{6}
                 +\frac{1}{2 \sqrt{3}} \cos 2\chi \sin 2 \phi; \hspace{0.4cm}
\end{eqnarray}
This is our generalistion of TBM (and trimaximal),
to all mixings with the $\nu_2$ tri-maximally mixed, 
which we refer to as `S3 group mixing'~\cite{char}.
Of course, the group matrix construction 
works in any flavour basis,
with the group matrices transformed appropriately
(ie.\ in any representation 
equivalent to the natural representation).

\section{Class Operators: Mutativity and Democracy}
An alternative way to generalise the group matrix construction
from C3 to S3, is to note that, with C3 being an abelian group, 
there is no distinction between the group elements 
and the group classes in that case. 
We showed~\cite{char} that exact tri/bi-maximal mixing
results by assuming that the charged-lepton mass matrix in the neutrino mass basis
(where the neutrino mass matrix is diagonal)
is a general S3 class matrix in the natural representation of the S3 class algebra
(leading us to an intriguing relation between the tri/bi-maximal mixing matrix
and the S3 group character table). 
We will not discuss this construction in any further detail here.

    Alerted to the relevance of class operators however, 
we went on to show~\cite{char} that in exact tri/bi-maximal mixing,
the neutrino mass matrix in the flavour basis 
is a class operator 
for the S1 $\subset$ S2 $\subset$ S3 canonical subroup chain~\cite{chen}
in the natural representation of the S3 group.
The individual class operators may be written:
\begin{eqnarray}
C(1) = I = \left( 
\matrix{ 1 & 0 & 0 \cr
         0 & 1 & 0 \cr
         0 & 0 & 1 } \right) \hspace{15mm}
C(2) = P(\mu \tau)  = \left( 
\matrix{ 1 & 0 & 0 \cr
         0 & 0 & 1 \cr
         0 & 1 & 0 } \right) \\ 
C(3) = P(e \mu) + P(\mu \tau)+P(\tau e)   
= \left( 
\matrix{ 1 & 1 & 1 \cr
         1 & 1 & 1 \cr
         1 & 1 & 1 } \right). \hspace{26mm}
\end{eqnarray}
The S2 class operator $C(2)$ is taken to be the mu-tau exchange operator
(`the mutau-tivity operator')\cite{muta}\cite{grim}
 having eigenvalues $M_i= \pm 1$.
The S3 class operator $C(3)$ might reasonably be called
the `democracy' operator, having eigenvalues $D_i=0,3$.
Together these two quantum numbers are sufficient to distinguish
the neutrino mass eigenstates in exact tri/bi-maximal mixing
(the identity operator $C(1)$ gives unity always
and could be thought of, 
eg.\ as just the modulus of the lepton number $I_i=|L_i|=1$).

The above operators commute
with each other
and are simutaneously diagonalisable (uniquely) 
by the tri/bi-maximal mixing matrix: 
\begin{eqnarray}
     \matrix{  \hspace{0.4cm} \nu_1 \hspace{0.4cm}
               & \hspace{0.2cm} \nu_2 \hspace{0.3cm}
               & \hspace{0.2cm} \nu_3  \hspace{0.4cm} }
                                      \hspace{00.2cm} \nonumber \\
     \matrix{  \hspace{0.4cm} (1,0) \hspace{0.0cm}
               & \hspace{0.cm} (1,3) \hspace{0.cm}
               & \hspace{0.0cm} (-1,0) \hspace{0.2cm} }
                                      \hspace{0.2cm} \nonumber \\
\hspace{1.0cm}U \hspace{0.3cm} = \hspace{0.3cm}
\matrix{ e \hspace{0.2cm} \cr
         \mu \hspace{0.2cm} \cr
         \tau \hspace{0.2cm} }
\left( \matrix{ \sqrt{\frac{2}{3}}  &
                     \frac{1}{\sqrt{3}} &
                                   0  \cr
 -\frac{1}{\sqrt{6}}  &
          \frac{1}{\sqrt{3}} &
 -\frac{1}{\sqrt{2}}   \cr
      \hspace{2mm}
 -\frac{1}{\sqrt{6}} \hspace{2mm} &
         \hspace{2mm}
            \frac{1}{\sqrt{3}} \hspace{2mm} &
                                 \frac{1}{\sqrt{2}}
                                       \hspace{2mm} \cr } \right).
\label{tb2}
\end{eqnarray}
Both mutativity ($M_i$) and democracy ($D_i$) 
are good quantum numbers (for neutrinos)
in exact tri/bi-maximal mixing, and 
in Eq.~\ref{tb2} they are used to label ($M_i,D_i$)
the columns of the mixing matrix, 
equivalently to the $\nu_i$ labels immeduately above. 

The neutrino mass-matrix 
in the charged-lepton flavour basis may now be written: 
\begin{equation}
M^2_{\nu}  =  sC(1)+tC(2)+uC(3). \label{s3s2}
\end{equation}
where $s$, $t$, $u$ are real.
Explicitly:
\begin{equation}
M^2_{\nu}  =  \left( \matrix{
s+t+u & u & u \cr
u   & s+u & t+u \cr
u & t+u & s+u } \right). \label{m3s2}
\end{equation}
The eigenvectors of Eq.~\ref{m3s2}
appear as the columns of Eq.~\ref{tb2}.
The eigenvalues 
are (the squares of) the neutrino masses:
\begin{eqnarray}
m_1^2 & = & s+t \\
m_2^2 & = & s+t+3u \\
m_3^2 & = & s-t.
\end{eqnarray}
The coefficients
$0 \le 3u \le -2t \le 2s$
(for $m_1^2 \le m_2^2 \le m_3^2$) 
are given in terms of
the neutrino masses by:
\begin{eqnarray}
s & = & \frac{m_1^2+m_3^2}{2} \\
t & = & \frac{m_1^2-m_3^2}{2} \\
u & = & \frac{m_2^2-m_1^2}{3}.
\end{eqnarray}

Using the mutautivity, democracy 
(and identity) quantum numbers,
we have the single mass formula:
\begin{equation}
m_i^2=sI_i+tM_i+uD_i
\end{equation}
Of course this is entirely equivalent to Eqs.~15-17, 
so in fact the fundamental gain in conceptual simplicity, 
coming from this last step, is at best a relatively modest one. 

\section{Conclusion}
The tri/bi-maximal hypothesis
is alive and well following the SNO/KAMLAND results,
giving an (at least approximate) description of lepton mixing.
For any mixing with the $\nu_2$ trimaximally mixed
the neutrino mass-matrix in the flavour basis satisfies 
an S3 invariant constraint (the `magic square' constraint).
In exact tri/bi-maximal mixing the mass matrix
is an S1 $\subset$ S2 $\subset$ S3 class opertaor,
whereby the neutrino mass eigenstates are distinguished
by their mutativity and democracy quantum numbers.

\section{Acknowledgements}
This work was supported by the UK 
Particle Physics and Astronomy Research Council (PPARC).
One of us (PFH) acknowledges the hospitality 
of the Centre for Fundamental Physics (CFFP)
at CCLRC Rutherford Appleton Laboratory.

\newpage

\noindent {\bf Appendix} \\

The plot reproduced below
is Figure~3 from our 1996 paper~\cite{hps2}  
showing the \mbox{`$5/9-1/3-5/9$'} LMA bathtub,
in the context of trimaximal mixing.
As shown in Section~2 above,
the same curve actually fits  
the present-day solar data rather well,
lending support to our general approach and
suggesting that we may have been on the right track.
In the LMA (large angle MSW~\cite{msw}) solution
to the solar neutrino problem,
matter effects override the vacuum mixing 
within the range defining the `bathtub', see Figure~3.
\begin{figure}[h]
\begin{center}
\epsfig{figure=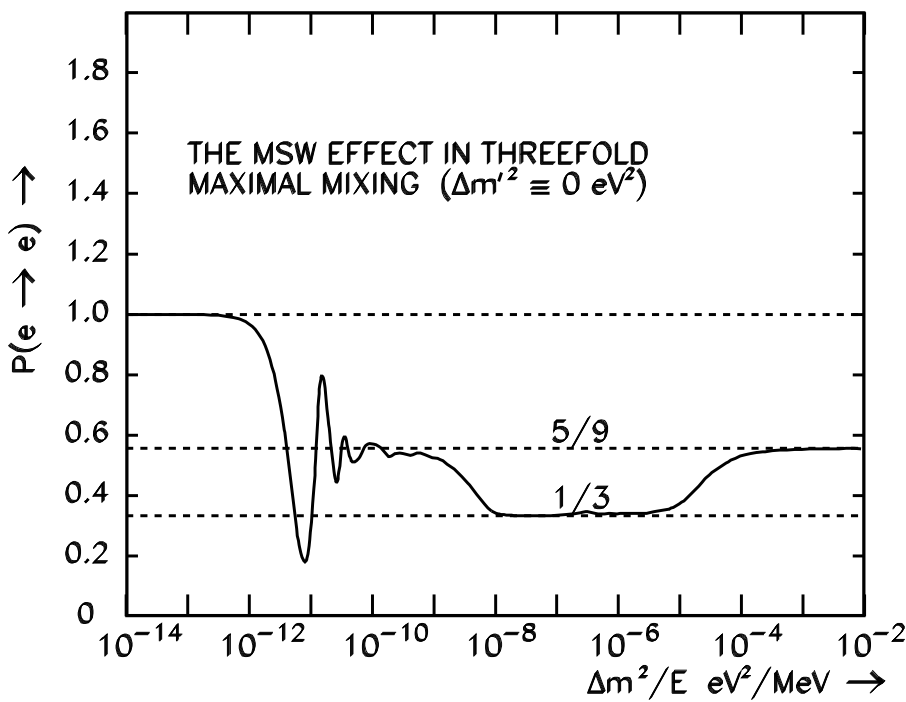,width=12.0cm}
\end{center}
\caption{The `$5/9-1/3-5/9$' bathtub in trimaximal mixing
from our 1996 paper.~\protect\cite{hps2}. 
Within the bathtub region one expects $P(e \rightarrow e)=1/3$,
while outside the bathtub we predicted the vacuum value 
$<P(e \rightarrow e)>$ $= |U_{e1}|^2+|U_{e2}|^2+|U_{e3}|^2=5/9$.
These predictions remain vaild within tri/bi-maximal mixing, 
because the solar neutrino is likewise 
trimaximally mixed in that case.\protect}
\label{fig:fivn}
\end{figure}
Given the value of the solar core density, 
at very {\em low} neutrino energies 
($\Delta m^2/E \sim 10^{-4}-10^{-5}$ eV$^2$/MeV) 
vacuum mixing dominates once again over matter effects
and the survival probability returns to its vacuum value
$<P(e \rightarrow e)>$ $= |U_{e1}|^2+|U_{e2}|^2+|U_{e3}|^2 =5/9$.
Given the rate of fall-off of the solar density with radius,
at {\em high} neutrino energies 
($\Delta m^2/E \simeq 10^{-9}-10^{-8}$ eV$^2$/ MeV)
`level jumping' occurs due to breakdown of the adiabaticity,
and we again recover the vacuum prediction.
Within the range of the bathtub (Figure~3)
the experiments simply measure the $\nu_e$ content
of the mass-eigenstate prepared by the Sun,
giving $P(e \rightarrow e) = 1/3$ there.
These predictions remain valid in tri/bi-maximal mixing, 
because the solar neutrino remains trimaximally mixed in that case.

\end{document}